\documentclass[%
reprint,
superscriptaddress,
showkeys,
amsmath,amssymb,
aps,
pra,
longbibliography,
floatfix,
]{revtex4-2}

\usepackage{defs}

\begin{document}

\title{Rydberg Atomic Antenna in Strongly Driven Multi-Electron Atoms}

\author{Stefanos Carlström\,\orcidlink{0000-0002-1230-4496}}%
\email{stefanos@mbi-berlin.de}
\email{stefanos.carlstrom@matfys.lth.se}
\affiliation{Max-Born-Institut, Max-Born-Straße 2A, 12489 Berlin, Germany}
\affiliation{Department of Physics, Lund University, Box 118, SE-221 00 Lund, Sweden}

\author{Jan Marcus Dahlström\,\orcidlink{0000-0002-5274-1009}}%
\affiliation{Department of Physics, Lund University, Box 118, SE-221 00 Lund, Sweden}

\author{Misha Yu Ivanov\,\orcidlink{0000-0002-8817-2469}}%
\affiliation{Max-Born-Institut, Max-Born-Straße 2A, 12489 Berlin, Germany}
\affiliation{Department of Physics, Imperial College London, South Kensington Campus, SW72AZ London, United Kingdom}
\affiliation{Institut für Physik, Humboldt-Universität zu Berlin, Newtonstraße 15, 12487 Berlin, Germany}

\author{Serguei Patchkovskii}%
\affiliation{Max-Born-Institut, Max-Born-Straße 2A, 12489 Berlin, Germany}

\date{\today}

\begin{abstract}
  We study the role of intermediate excitations of Rydberg states as
  an example of Kuchiev's \enquote{atomic antenna} in above-threshold
  ionization of xenon, in particular their effect on the coherence
  between the spin--orbit-split states of the ion. We focus on the case
  of a laser frequency close to resonant with the spin--orbit
  splitting, where a symmetry (parity) argument would preclude any
  coherence being directly generated by strong-field ionization. Using
  \emph{ab initio} simulations of coupled multielectron spin--orbit
  dynamics in strong laser fields, we show how field-driven
  rescattering of the trapped Rydberg electrons introduces efficient
  coupling between the spin--orbit-split channels, leading to
  substantial coherences, exceeding \SI{10}{\percent} for some photon
  energies.
\end{abstract}

\keywords{Ultrafast spin--orbit interaction, above-threshold
  ionization, Freeman resonances, atomic antenna, rescattering}
\maketitle

\section{Introduction}
Spin--orbit effects are usually neglected in the interaction with
strong infrared (IR) fields. Few exceptions include the development of
consistent treatment within the \(R\)-matrix method \cite{Wragg2019},
recent experiments \cite{Kuebel2019, Goulielmakis2010} on imaging the
spin--orbit breathing of a hole created by strong-field ionization, and
the generation of spin-polarized photoelectrons \cite{Hartung2016}
following the proposal of \cite{Barth2013,Barth2014}. In all these
cases \cite{Kuebel2019,
  Goulielmakis2010,Hartung2016,Barth2013,Barth2014}, the dipole
approximation holds, ensuring that no transitions between the
spin--orbit-split states were induced by the incident IR field.

In this article we show that, even in the dipole
approximation, strong IR fields trigger transitions between the
spin--orbit-split states of the ion via a mechanism resembling the
\enquote{atomic antenna} of \textcite{Kuchiev1987}. In our case, an
active electron driven by strong IR field is trapped into a long-lived
Rydberg orbit.  Oscillating in the IR field, it transfers the energy
to the core via non-dipole electron--electron interaction. This atomic
antenna breaks the dynamic symmetry with respect to the polarization
of the linearly polarized driving laser field \cite{Tzur2021}. It
thereby induces coherence between the spin--orbit-split states of the
ion, reduces the entanglement between the ion and the photoelectron,
and manifests itself in the photoelectron spectra.

This article is arranged as follows: in section~\ref{sec:theory}, we
introduce the degree of coherence between ionic states, and discuss
why parity-conservation arguments require the coherence to vanish,
when the photon energy matches the spin--orbit splitting; in
section~\ref{sec:results}, we present the main computational results
that contradict these expectations, as well as our
explanation. Finally, section~\ref{sec:conclusions} concludes the
article.

\section{Theory}
\label{sec:theory}
We consider a xenon atom, initially in the ground state, interacting
with an IR pulse with carrier photon energy \(\hbar\omega\) close to the
spin--orbit splitting of the cation,
\(\DEso\approx\SI{1.3}{\electronvolt}\),
\(\eta\defd \hbar\omega/\DEso\sim 1\) (atomic units
\(\hbar=e=a_0=\electronmass=1\) are used in the following). Our
calculations include all relevant electronic excitations (i.e.\ single
excitations/ionizations from \orbitalc{}, \orbitalb{}, or \orbitala{}
are allowed), and account for spin--orbit coupling effects. We solve
the time-dependent Schrödinger equation in the dipole approximation
and the length gauge, for a configuration-interaction singles
\emph{Ansatz} that allows single excitation/ionization from a
Hartree--Fock (HF) reference \cite{Nesbet1955, Loewdin1955a,
  Krause2005, Rohringer2006, Greenman2010PRA, Carlstroem2022tdcisI,
  *Carlstroem2022tdcisII}. The spin--orbit interaction is treated using
an energy-consistent relativistic effective-core potential
\cite{Peterson2003} (see \cite{Zapata2022} for an alternative option,
based on the four-component Dirac equation). Ion-resolved
above-threshold ionization (ATI) photoelectron spectra are computed
\cite{Carlstroem2022tdcisI, *Carlstroem2022tdcisII} using the tSURFF
\cite{Ermolaev1999, Ermolaev2000, Serov2001, Tao2012NJoP,
  Scrinzi2012NJoP} and iSURFV \cite{Morales2016-isurf}
techniques. From these spectra, we compute the reduced density matrix
\(\{\coherence{IJ}\}\), obtained by tracing over the photoelectron
degrees of freedom. We then form the normalized \emph{degree of
  coherence}:
\begin{equation}
  \label{eqn:degree-of-coherence}
  \degreeofcoherence{IJ} \defd
  \frac{\coherence{IJ}}{\sqrt{\coherence{II}\coherence{JJ}}},
\end{equation}
where \(\coherence{IJ}\) is the coherence between \(I\) and \(J\), and
\(\coherence{II}\) and \(\coherence{JJ}\) are the populations in the
ion states \(I\) and \(J\), respectively. Further details are given in
Appendix~\ref{app:methods}.

Let us first consider coherence between the spin--orbit-split states of
the ion generated by ionization \cite{Goulielmakis2010, Barth2014,
  Pabst2016, Ruberti2018, Ruberti2019a, Ruberti2021}. The final state
of the system \enquote{ion+photoelectron} is
\begin{equation}
  \label{eqn:ion-electron-state}
  \begin{aligned}
    \ket{\Psi}
    &=
      \pket*{I}{\chi_I}+\pket*{J}{\chi_J} \\
    &=
      (\ket*{I}+
      \ket*{J}
      \eloverlap)
      \ket*{\chi_I} +
      \ket*{J}
      (\ket*{\chi_J} -
      \eloverlap
      \ket*{\chi_I}),
  \end{aligned}
\end{equation}
where we choose \(\eloverlap\defd\braket*{\chi_I}{\chi_J}\) as a measure of
the factorizability of the wavefunction, and antisymmetrization with
respect to the coordinates of the photoelectron is implied. Coherent
spin--orbit dynamics in the ion requires non-zero overlap between the
continuum electron wavepackets correlated to the ionic states
\(\ket*{I}\) and \(\ket*{J}\), respectively: \(\eloverlap \neq
0\). Perfect overlap \(\abs{\eloverlap}=1\) corresponds to
\SI{100}{\percent} degree of coherence, since
\eqref{eqn:ion-electron-state} factorizes into
\begin{equation*}
  \ket{\Psi}=(\ket*{I}+\ce^{\im\phi}\ket*{J})\ket*{\chi},
\end{equation*}
for some phase \(\phi\). Perfect electron--ion entanglement corresponds to
\(\eloverlap=0\).

\begin{figure}[t]
  \centering
  \includegraphics{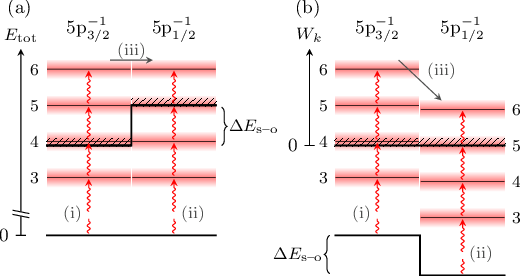}
  \caption{Sketch of the energy diagram for ATI from xenon for
    \(\engratio\defd\hbar\omega/\DEso=0.9\); the numbers indicate the amount of
    photons necessary to reach a certain final energy. The three most
    important pathways are: (i) direct ionization into the \channela{}
    channel including elastic rescattering, (ii) direct
    ionization/elastic rescattering in the \channelb{} channel, and
    (iii) indirect contributions due to inelastic rescattering from
    \channela{} to \channelb{} (this dominates over rescattering in
    the other direction). (a) Total energy (photoelectron + ion),
    relative to the field-free neutral atom. In this picture, the
    energy conservation in inelastic scattering is easily
    seen. (b) Energies of the photoelectrons in each
    channel. Non-zero photoelectron overlap is necessary for a
    coherence between the ion cores to exist. For \(\engratio\sim1\),
    photoelectrons of similar kinetic energies are due to absorption
    of a different number of photons.}
  \label{fig:energy-diagram}
\end{figure}

\begin{figure}[htb]
  \centering
  \includegraphics{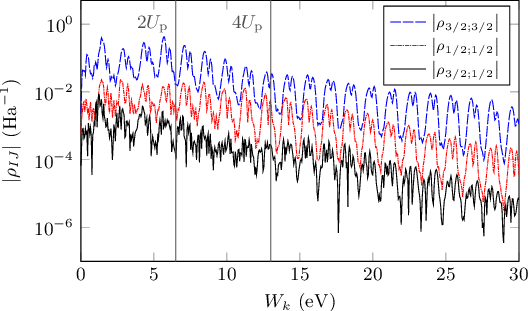}
  \caption{Calculated ATI spectrum of xenon using
    \SI{4.4e13}{\watt\per\centi\meter\squared} with
    \(\engratio\approx0.96\) and a pulse duration of
    \SI{30}{\femto\second}. The dashed blue line is the photoelectron
    spectrum \(\rho_{3/2;3/2}\) correlated with the \channela{}
    ionization channel; the dot-dashed red line the corresponding
    spectrum \(\rho_{1/2;1/2}\) for the \channelb{} channel; and the
    black solid line is the energy-resolved coherence
    \(\rho_{3/2;1/2}\) between the two channels. Above approximately
    \(4\ponderomotive\), the spectrum in \channelb{} and the coherence
    exhibit structures which are very similar to the spectrum in
    \channela{}; we infer that the \channelb{} channel is populated
    almost exclusively through rescattering in this energy region [see
    (iii) in Figure~\ref{fig:energy-diagram}].}
  \label{fig:ati-spectrum}
\end{figure}

By a symmetry argument, zero coherence and perfect entanglement are
expected for \(\engratio=1\): non-zero coherence and hence
\(\eloverlap\neq0\) requires that the two photoelectron wavepackets
overlap in energy. After absorption of \(q\) photons of energy
\(\omega\), the photoelectron energy is
\begin{equation}
  \label{eqn:kinetic-energy}
  \kineng = q\omega -
  \ionpotential[I] -
  \ponderomotive -
  \frac{\delta\alpha_I}{4} F^2
  = q\omega -
  \ionpotential[I] -
  \ponderomotive
  (1 + \omega^{2}\delta\alpha_I),
\end{equation}
where \(\omega\) is the driving laser frequency, \(\ionpotential[I]\) is
the ionization potential in ionization channel \(I\),
\(\ponderomotive=F^2/4\omega^2\) the \emph{ponderomotive potential} of the
electric field with peak amplitude \(F\), and \(\delta\alpha_I\) the difference
between the polarizabilities of the ground state of the neutral and
the state of the ion. For \(\engratio\sim 1\), the photoelectron peaks
correlated to the \channela{} and \channelb{} ion cores coincide in
energy when one extra photon is absorbed in the \channelb{} channel
(see Figure~\ref{fig:energy-diagram}). Thus, the photoelectron
associated with \(\channela\) would have opposite parity compared to
\(\channelb\), while the \channela{} and \channelb{} ion cores have
the same parity. The overall parity would thus be opposite between the
channels, implying \(\eloverlap=0\) by symmetry, precluding any
coherence. If very short, broadband pulses are used, a non-zero
coherence can nonetheless result, due to the energetic overlap of two
successive ATI peaks belonging to the two thresholds. This is the
mechanism behind the coherence observed by
\textcite{Goulielmakis2010}, who use pulses of \SI{3.8}{\femto\second}
duration. This coherence diminishes when longer pulses are used, and
is expected to disappear entirely for the much longer pulses
(\(\ge\SI{15}{\femto\second}\)) used in the present work.

\section{Results}
\label{sec:results}
We begin by considering ionization by a \SI{30}{\femto\second} pulse,
tuned just below the spin--orbit splitting (\(\eta=0.96\)). As can be seen
from the simulation results in Figure~\ref{fig:ati-spectrum}, there is
non-zero coherence between \channela{} and \channelb{}, where the ATI
peaks in the respective channels overlap energetically.

\begin{figure}[htb]
  \centering
  \includegraphics{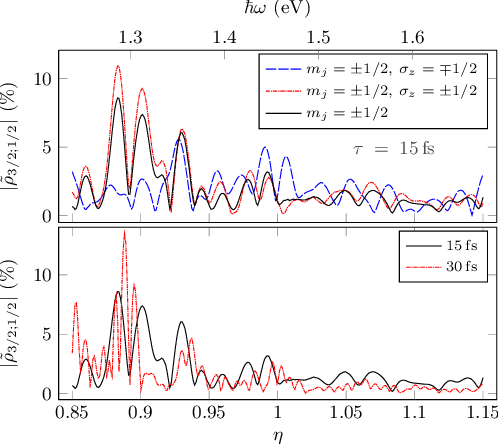}
  \caption{Ionic coherence in xenon. \emph{Upper panel}: the degree of
    coherence [Eq.~\eqref{eqn:degree-of-coherence}] between the
    ionization channels \channela{} and \channelb{} as a function of
    the ratio \(\engratio\) between the photon energy and the
    spin--orbit splitting, resolved on the \(m_j\) quantum number of
    the ion and the spin \(\spin\) of the photoelectron. Due to the
    cylindrical symmetry of the ionization process, there is a mirror
    symmetry in the combinations of \(m_j\) and
    \(\spin\). Additionally tracing out \(\spin\) leads to the final
    degrees of coherence for \(m_j=\pm1/2\), which coincide. \emph{Lower
      panel}: the degree of coherence for two different pulse
    durations; solid black line: \(\tau=\SI{15}{\femto\second}\),
    dot-dashed red line: \(\tau=\SI{30}{\femto\second}\).}
  \label{fig:degree-of-coherence}
\end{figure}
Figure~\ref{fig:degree-of-coherence} shows the calculated degree of
coherence \eqref{eqn:degree-of-coherence} between the \channela{} and
\channelb{} ion cores, as a function of \(\engratio\). Contrary to the
symmetry-based expectation, we see substantial coherence, even
exceeding \SI{10}{\percent} for some \(\engratio\). We trace its
origin to frustrated tunnelling \cite{Nubbemeyer2008, Eichmann2009,
  Zimmermann2017} --- trapping of the electron into Rydberg states after
optical tunnelling from the ground state. Once the neutral atom is
\enquote{parked} in the intermediate, excited state for an extended
amount of time, there is an opportunity to undergo multiple successive
(Stokes--)Raman transitions that each increase the system energy by a
small amount, while conserving the parity. After multiple such Raman
transitions have occured, the energy may increase enough to bridge the
energy gap to the next ATI order. Upon subsequent ionization and
rescattering into the other ion channel, the overlap of photoelectrons
of the same parity and energy explains the observed coherence.

In the frequency domain, frustrated tunnelling followed by ionization
corresponds to the so-called \emph{Freeman resonances}
\cite{Freeman1987} imprinted on top of the photoelectron peaks. The
apparent \enquote{parity violation} is a manifestation of the dynamic
symmetry being broken due to the simultaneous presence of the Freeman
resonances, and the spin--orbit interaction. The Freeman resonances
introduce memory in the time evolution, breaking time-reversal
symmetry, or equivalently, spatial inversion symmetry between the
response of the system to two successive half-cycles. Simultaneously,
the spin--orbit coupling leads to the mixing of the ionic spin--orbit
channels. Together, these two effects demote the photoelectron parity
from a selection rule to a \emph{propensity rule} \cite{Tzur2021}. We
stress that parity conservation of the \emph{whole} wavefunction may
\emph{not} be violated, whereas there is no such guarantee for the
constituent parts. That parity with respect to the \(\ell\) quantum
number of the photoelectron is only a propensity rule has also been
observed in an analogous example in single-photon spectroscopy of
xenon \cite{Dill1973, Samson1973}, where it has also been linked to
the interaction with the core electrons.

\begin{table}[bp]
  \caption{\label{tab:antenna-transitions} Some dipole-allowed
    transitions in xenon \cite{Saloman2004} which are likely
    candidates for the \enquote{antenna transition}, given their
    energies and compositions.}
  \begin{ruledtabular}
    \begin{tabular}{lllll}
      \Bstrut \unitlabel{\(\Delta E_{ki}\)}{\electronvolt} & \(i\) Conf. & Term & \(k\) Conf. & Term \\
      \hline
      \Tstrut
      \num{1.353} & \(\conf{5p^5(\term*{2}{P}[3/2])6s}\) & \term*{2}{[3/2]}[1] & \(\conf{5p^5(\term*{2}{P}[3/2])6p}\) & \term{2}{[3/2]}[1] \\
      \Tstrut
      \num{1.332} & \(\conf{5p^5(\term*{2}{P}[1/2])6s}\) & \term*{2}{[1/2]}[1] & \(\conf{5p^5(\term*{2}{P}[3/2])7p}\) & \term{2}{[1/2]}[1] \\
      \Tstrut
      \num{1.265} & \(\conf{5p^5(\term*{2}{P}[3/2])6s}\) & \term*{2}{[3/2]}[2] & \(\conf{5p^5(\term*{2}{P}[3/2])6p}\) & \term{2}{[1/2]}[1] \\
      \Tstrut
      \num{1.249} & \(\conf{5p^5(\term*{2}{P}[3/2])6s}\) & \term*{2}{[3/2]}[1] & \(\conf{5p^5(\term*{2}{P}[3/2])6p}\) & \term{2}{[5/2]}[2] \\
    \end{tabular}
  \end{ruledtabular}
\end{table} The atomic antenna by
\textcite{Kuchiev1987} lends a complementary perspective: the
intermediate excited Rydberg states of the neutral are in some aspects
very similar to free electrons. A resonance structure is built up in
the (Stark-shifted) quasi-continuum of the Rydberg states, that
similarly to an antenna can be used to channel energy into the system
and thereby drive transitions in the ion core. It is of course
necessary that the antenna is \enquote{sensitive} to the radiation
\(\hbar\omega\) impinging on it, such that it may efficiently couple the energy
into the system; this is the case if a pair of Rydberg states is
separated by \(\hbar\omega\). Furthermore, one or both of the states involved
in the transition must bridge the ion manifold, i.e.\ have components
in both the \channela{} and \channelb{} manifolds. A few of the likely
candidates for the antenna transitions are listed in
Table~\ref{tab:antenna-transitions}. This is the frequency-domain
perspective of the inelastic rescattering. To confirm the antenna
picture, we have investigated transitions for which
\(\engratio\in[0.85,1.15]\) and their strengths. The details are given
in Appendix~\ref{app:confirming-antenna}, along with alternative
explanations that we have considered, such as depletion, envelope
effects, and single-state coherence.

\begin{figure}[t]
  \centering
  \includegraphics{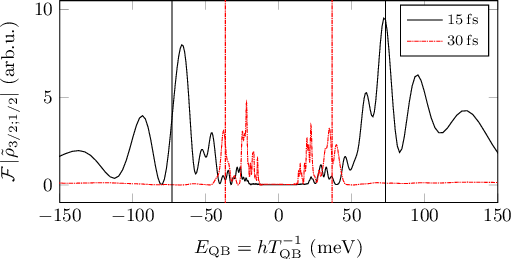}
  \caption{Fourier transform of the degree of coherence depicted in
    Figure~\ref{fig:degree-of-coherence}, for the two different pulse
    durations; solid black line: \(\tau=\SI{15}{\femto\second}\),
    dot-dashed red line: \(\tau=\SI{30}{\femto\second}\). Since
    \(\hbar\omega\equiv\engratio\DEso\) has dimension of energy, the conjugate
    variable has dimension of time. Here, we instead plot the peaks as
    a function of the quantum beat \emph{energy} \(\EQB\), i.e.\ the
    energy separation between neighbouring antenna transitions. The
    vertical lines indicate the spectral width of the driving pulse,
    \(\dE = \hbar\sqrt{4\ln2}\tau^{-1}\), which for
    \(\tau=\SI{15}{\femto\second}\) is
    \(\sim\SI{73.1}{\milli\electronvolt}\) and
    \(\sim\SI{36.5}{\milli\electronvolt}\) for
    \(\tau=\SI{30}{\femto\second}\).}
  \label{fig:coherence-fringes}
\end{figure}
To further investigate the role of the Rydberg states excited via the
Freeman resonances, we perform a Fourier transform of the degree of
coherence along the \(\eta\) axis. This analysis reveals quantum beat
periods of the excited wavepacket, which constitute a fingerprint of
the atomic antenna. By inverting the quantum beat periods, we instead
get the energy separation between neighbouring antenna transitions,
which is shown in Figure~\ref{fig:coherence-fringes}. As is evident
from Figure~\ref{fig:coherence-fringes}, the very complex coherence
patterns in Figure~\ref{fig:degree-of-coherence} is in fact due to a
small number of individual antenna transitions. These transitions
occur for energy separations close to the bandwidth of the driving
pulse. This is not an accident: transitions at these energies reach an
optimal balance between the available photon fluence (decreasing away
from the carrier frequency, making the transition less likely) and the
number of photons needed to be emitted/absorbed to close the spin--orbit
gap, which decreases for larger energy separations (increasing the
transition probability). When the pulse duration increases, the
spectral bandwidth decreases. This imposes stricter requirements on
which antenna transitions are in resonance with the driving field with
photon energy \(\hbar\omega\). It is more likely to find such transitions among
the higher-lying states in the Rydberg quasi-continuum, which are more
closely spaced energetically. This explains why for the longer pulse
duration, we in Figure~\ref{fig:coherence-fringes} observe quantum
beat components of the wave packet with comparatively smaller
\(\EQB\), corresponding to the more tightly spaced peaks in
Figure~\ref{fig:degree-of-coherence}.

This circumstance also helps us understand why the longer pulse
duration can produce \emph{larger} degrees of coherence, when we might
have expected the opposite; decreasing spectral bandwidth leads to
narrower photoelectron peaks \cite{Petite1987}, which in turn leads to
smaller energetic overlap between the ATI progressions. However, as
long we are in resonance with the antenna, a longer pulse is
beneficial since we can transfer population into the different
pathways, while maintaining coherence. This is reminiscent of the
previously studied case of weak-photon ionization of xenon
\cite{Carlstroem2018}, where longer pulse durations also led to
increased ionic coherence, albeit for a simpler resonance condition.

We emphasize that, although the electron--electron interaction is
crucial for the effectiveness of the antenna mechanism, the initial
asymmetry is created by the laser field, which imposes the natural
quantization axis on the system. The asymmetry is then transferred to
the electron spin through the spin--orbit interaction (electron spins
do not couple to the laser field in the dipole
approximation). Finally, the electron--electron interaction provides
the very efficient coupling between the Rydberg electron (the
\enquote{antenna}) and the ion core. Thus, all three interactions are
essential, with each playing a distinct role in the process.

\section{Conclusions}
\label{sec:conclusions}

We have shown that through the intermediate Rydberg state dynamics, we
can introduce coherence between ionization pathways that would
otherwise have opposite parity by symmetry. The coherence is sensitive
to the frequency and duration of the ionizing laser pulse, and allows
us to identify the effect of the Rydberg atomic antenna essentially
background-free.

\begin{acknowledgements}
  We would like to acknowledge the help of the late Oleg Zatsarinny in
  estimating the viability of this project. The work of SCM has been
  supported through scholarship 185-608 from \emph{Olle Engkvists
    Stiftelse}. JMD acknowledges support from the \emph{Knut and Alice
    Wallenberg Foundation} (2017.0104 and 2019.0154), the
  \emph{Swedish Research Council} (2018-03845) and \emph{Olle
    Engkvists Stiftelse} (194-0734). MI acknowledges support from
  \emph{Deutsche Forschungsgemeinschaft} (IV~152/10-1).
\end{acknowledgements}

\appendix

\section{Methods}
\label{app:methods}

We employ Hartree atomic units and implied summation/integration over
indices, orbitals, momenta, and/or spins appearing on only one side of
an equation.

\subsection{Grid-Based Time-Dependent Configuration-Interaction
  Singles}
The derivation of the equations of motion (EOMs), and a detailed
description of the propagator are given in \cite{Carlstroem2022tdcisI,
  *Carlstroem2022tdcisII}; the EOMs agree with those of
\textcite{Rohringer2006,Greenman2010PRA}, apart from the fact that
spin-restriction is \emph{not} imposed in the present work, i.e.\ we
are solving the two-component Schrödinger equation.

The TD-CIS EOMs describe the time evolution of the amplitude \(c_0\)
for the Hartree–Fock (HF) reference state, and the particle orbital
\(\contket*{k}\) emanating from the occupied (time-independent)
orbital \(\ket{k}\). The different particle--hole channels can couple
via either the laser interaction or the Coulomb interaction:
\begin{equation}
  \label{eqn:td-cis-eoms-simplified}
  \begin{aligned}
    \imdt c_0
    &=
    {\matrixel*{k}{\laserinteraction}{\cont{k}}}, \\
    \imdt\contket*{k}
    &=
    (-\orbitalenergy{k}+{\fock})\contket*{k} +
    {c_0\laserinteraction\ket{k}} -
    {\matrixel{l}{\laserinteraction}{k}}\contket*{l} \\
    &
    \hphantom{=}
    -{(\direct[lk]-\exchange[lk])\contket*{l}} -
    \lagrange{\cont{k}i}\ket{i},
  \end{aligned}
\end{equation}
where \(\orbitalenergy{k}\) is the field-free energy of the occupied
orbital \(\ket{k}\),
\((\fock-\laserinteraction-\orbitalenergy{k})\ket{k}=0\), the Fock
operator is defined as
\(\fock\defd \hamiltonian + \direct[ii] - \exchange[ii]\), with the
one-body Hamiltonian containing the interaction with the external
laser field,
\(\hamiltonian \defd p^2/2 + \nuclearpotential(\vec{r}) +
\laserinteraction\),
\(\laserinteraction\defd \fieldamplitude{t}\cdot\vec{r}\), and the
\emph{direct} and \emph{exchange interaction} potentials are given by
their action on an orbital
\begin{equation*}
  \begin{aligned}
    \direct[cd]\ket{e} &\defd
    \orbital{e}(\spatialspin_1)
    \int\frac{\diff{\spatialspin_2}}{\abs{\vec{r}_1-\vec{r}_2}}
    \conj{\orbital{c}}(\spatialspin_2)
    \orbital{d}(\spatialspin_2), \\
    \exchange[cd]\ket{e} &\defd
    \orbital{d}(\spatialspin_1)
    \int\frac{\diff{\spatialspin_2}}{\abs{\vec{r}_1-\vec{r}_2}}
    \conj{\orbital{c}}(\spatialspin_2)
    \orbital{e}(\spatialspin_2)
    \equiv \direct[ce]\ket{d},
  \end{aligned}
\end{equation*}
where \(\spatialspin_{1,2}\) refer to both spatial and spin
coordinates of the orbitals. As we consider atoms in the present work,
the particle orbitals \(\contket*{k}\contket*{l}...\) are conveniently
expanded in a tensor product basis formed from spinor spherical
harmonics \citep[i.e.\ \(n\ell jm_j\);\ see \S 7.2
of][]{Varshalovich1988} and finite-differences for the radial
dimension \cite{Adler1984, Krause1999TJoPCA}. Finally, the Lagrange
multiplier \(\lagrange{\cont{k}i}\) ensures that \(\contket*{k}\) at
all times remains orthogonal to the occupied orbital \(\ket{i}\).

Because we are working in the dipole approximation,
\(\laserinteraction\) includes \electricdipole{} transitions only. As
discussed in the main text, \channela{} and \channelb{} have the same
parity, which means
\(\matrixel*{\channela}{\laserinteraction}{\channelb}=0\). However,
even if the dipole-forbidden (\electricquadrupole{} and
\magneticdipole{}) transitions between \channela{} and \channelb{}
were to be included, they would be so minuscule
\cite{Garstang1964,Nandy2015} that the resulting coherence would be
\(\sim\num{e-7}\) to \(\sim\num{e-9}\). Instead, in our simulations we find
coherence \(\sim\num{e-2}\) for \emph{all} \(\engratio\sim1\).

\subsection{Atomic Structure and Pulse Parameters}
The EOMs \eqref{eqn:td-cis-eoms-simplified} as formulated would yield
the same result as a one-component calculation, i.e.\ there would be
no effect due to the spin of the electrons. To implement spin--orbit
coupling (as well as corrections due to scalar-relativistic effects),
and at the same time reducing the number of electrons we need to treat
in the calculation, we replace the scalar potential
\(\nuclearpotential\) by the relativistic effective core potential
(RECP) of \textcite{Peterson2003}, which models the nucleus and the
\conf{1s}--\conf{3d} electrons according to
\begin{equation*}
\pseudopotential(\vec{r}) =
  -\frac{Q}{r} +
  B_{\ell j}^k \exp(-\beta_{\ell j}^k r^2)
  \proj[\ell j],
\end{equation*}
where \(Q=26\) is the residual charge, \(\proj[\ell j]\) is a projector
on the spin--angular symmetry \(\ell j\), and \(B_{\ell j}^k\) and
\(\beta_{\ell j}^k\) are numeric coefficients found by fitting to
multiconfigurational Dirac--Fock all-electron calculations of the
excited spectrum. For a thorough introduction to RECPs, see e.g.\ the
review by \textcite{Dolg2011}.

The radial grid consists of 527 points extending to \SI{90.4}{Bohr}
with the spacing smoothly varying according to
\cite{Krause1999TJoPCA}:
\begin{equation*}
  r_j = r_{j-1} + \rho_{\textrm{min}} + (1 - \ce^{-\alpha
    r_{j-1}})(\rho_{\textrm{max}}-\rho_{\textrm{min}}),
\end{equation*}
with \(r_1=\rho_{\textrm{min}}/2\),
\(\rho_{\textrm{min}}=\SI{0.1154}{Bohr}\),
\(\rho_{\textrm{max}}=\SI{0.1768}{Bohr}\), and \(\alpha=0.3\). The
spin--angular grid is limited to \(\Delta m_j=0\) since we only consider
linearly polarized light. For pulses of duration
\SI{15}{\femto\second} we use \(\ellmax=40\), and for
\SI{30}{\femto\second} \(\ellmax=60\).

Finally, since the calculation is performed in a finite computational
domain, we use Manolopoulos'~\cite{Manolopoulos2002}
transmission-free \emph{complex-absorbing potential} covering the last
\SI{12.57}{Bohr} at the far end of the box, with a design parameter
\(\delta\approx0.21\); this choice gives \(<\SI{1}{\percent}\) reflection for
photoelectrons with kinetic energies above \SI{3.4}{\electronvolt}
(\(\iff k_{\textrm{min}}=\SI{0.5}{au}\)).

\begin{table}[!h]
  \caption{\label{tab:xenon-energies} Calculated ionization potentials
    of the 5\{s,p\} electrons of xenon, compared with their
    experimental values. The corresponding Keldysh parameters, for the
    range of photon energies used, indicate that ionization is in a
    regime intermediate between the multi-photon and tunneling
    limits.}
  \begin{ruledtabular}
    \begin{tabular}{lrrrr}
      Hole & \unitlabel{\(\ionpotential\)}{\electronvolt} & Exp.\ \cite{Saloman2004} \unitbracket{\electronvolt} & \unitlabel{\(\Delta\)}{\electronvolt}
      & Keldysh \(\keldysh\)\Bstrut\\
      \hline
      \Tstrut
      \channelc & \(27.927\) & \(23.397\) & \(4.530\) & \(1.84\)--\(2.25\)\\
      \Tstrut
      \channelb & \(13.483\) & \(13.436\) & \(0.047\) & \(1.28\)--\(1.73\)\\
      \Tstrut
      \channela & \(12.026\) & \(12.130\) & \(-0.104\) & \(1.21\)--\(1.63\)\\
    \end{tabular}
  \end{ruledtabular}
\end{table}
With these grid parameters, the ionization potentials for the xenon
model (only 5s and 5p orbitals are allowed to ionize) are given in
Table~\ref{tab:xenon-energies}; the calculated spin--orbit splitting is
approximately \(\DEso\approx\SI{1.46}{\electronvolt}\). The deviation from
the experimental ionization potential is much larger for \channelc{};
this is to be expected at the CIS level of theory, where the ion is
not allowed to relax. This is however immaterial for the present work,
since its ionization fraction is negligible.

The driving field frequency is scanned across the range
\(\engratio\defd\hbar\omega/\DEso\in[0.85,1.15]\) \(\implies\)
\(\hbar\omega\) \(=\) \SIrange{1.24}{1.68}{\electronvolt}, and its intensity
\(I_0=\SI{4.4e13}{\watt\per\centi\meter\squared}\) \(\implies\)
\(\ponderomotive\) \(=\) \SIrange{4.12}{2.25}{\electronvolt} is chosen
such that the ionization remains at the level of a few percent. The
pulse duration is \SI{15}{\femto\second} or \SI{30}{\femto\second},
and the pulse shape is a smoothly truncated Gaussian
\cite{Patchkovskii2016}, with \(t_1=\SI{25.5}{\femto\second}\),
\(t_2=\SI{38.2}{\femto\second}\) and \(t_1=\SI{51.0}{\femto\second}\),
\(t_2=\SI{76.4}{\femto\second}\), respectively. The time propagator is
second-order accurate, and 2000 steps per carrier cycle are taken,
which yields a time step \(\timestep\) varying from
\SIrange{1.67}{1.23}{\atto\second}, for the range of values of
\(\engratio\) quoted above.

\subsection{Photoelectron Spectra and Ion Coherences}

Photoelectron spectra are computed using a multichannel extension
\cite{Carlstroem2022tdcisI, *Carlstroem2022tdcisII} of the tSURFF
\cite{Ermolaev1999, Ermolaev2000, Serov2001, Tao2012NJoP,
  Scrinzi2012NJoP} and iSURFV \cite{Morales2016-isurf} techniques,
yielding the familiar close-coupling \cite{Fritsch1991}
decomposition of the wavefunction, resolved on final ion state \(I\),
and photoelectron momentum \(\vec{k}\) and spin \(\spin\) (it is
assumed that the ion and photoelectron sufficiently separated, such
that antisymmetrization can be safely omitted):
\[\ket{\Psi} = c_{I\vec{k}\spin}\ket{I}\ket{\vec{k}\spin}.\]
From this long-range \emph{Ansatz}, we can form the density matrix of
the total system
\begin{equation*}
  \densitymatrix_{I\vec{k}'\spin';J\vec{k}''\spin''} \defd \ketbra{\Psi}{\Psi} =
  c_{I\vec{k}'\spin'}
  \ket{\vec{k}'\spin'}
  \ketbra{I}{J}
  \bra{\vec{k}''\spin''}
  \conj{c_{J\vec{k}''\spin''}},
\end{equation*}
and by subsequently tracing out the photoelectron, the \emph{reduced
  density matrix}, expressing the coherence between ion states
\begin{equation*}
  \coherence{IJ} =
  \bra{\vec{k}\spin}
  \matrixel{I}{\densitymatrix}{J}
  \ket{\vec{k}\spin} =
  c_{I\vec{k}\spin}
  \conj{c_{J\vec{k}\spin}},
\end{equation*}
(the population for the ion state \(I\) is
\(\coherence{I}\equiv\coherence{II}\)). These quantities are used to
compute the degree of coherence as shown in
Eq.~\eqref{eqn:degree-of-coherence}.

\section{Confirming the Atomic Antenna}
\label{app:confirming-antenna}
Below, we will discuss various aspects of the atomic antenna
\cite{Kuchiev1987}, and avenues we have pursued to confirm that this
proposed mechanism is indeed responsible for the observed symmetry
breaking and non-vanishing coherence.

\subsection{Influence of Depletion}

Since the degree of coherence is on the order of a few percent,
similar to the level of ionization for the intensity chosen, an
alternative explanation could be depletion-induced residual
coherence. This would be a memory effect, similar to hole-burning,
deviating from the cycle-to-cycle adiabaticity and breaking the
time-translation symmetry \cite{Jankowiak1993,Goll2006}. To rule out
this possibility, we artificially prevented the depletion of the
ground state by renormalizing the ground state amplitude after every
time step, which did not appreciably change the final coherence.

\subsection{Dynamical Effects due to the Envelope}

We also investigated whether the dynamical AC Stark shift of the
Rydberg states due to the envelope of the laser field had any
influence on the coherence. Substituting the Gaussian envelope by a
flattop pulse, removes most of the dynamical shifts, leaving only a
constant AC Stark shift. The degree of coherence was mostly unaffected
by this change, only increasing by a few percent.

\subsection{Removing one Rydberg State}

\begin{figure}[b]
  \centering
  \includegraphics{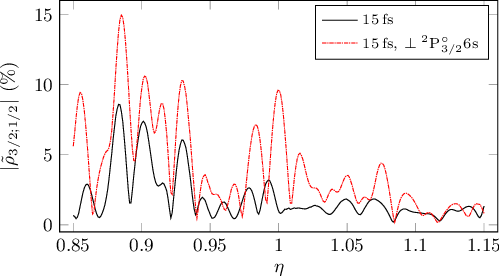}
  \caption{Effect on the degree of coherence by removing
    \intermediatestate{} from the calculation, see
    Equation~\eqref{eqn:projected-propagator}; black, solid line, the
    degree of coherence for a \SI{15}{\femto\second} pulse (same as
    seen in Figure~\ref{fig:degree-of-coherence} of the main article),
    and red, dot-dashed line, the degree of coherence for the same
    pulse, but with \intermediatestate{} projected out.}
  \label{fig:coherence-rydberg-removed}
\end{figure}

We next consider the effects of specific Rydberg states; we begin by
confirming that the Rydberg states, populated via frustrated
tunnelling, are important in the formation of the antenna. To test
this hypothesis we repeated the calculation, while preventing the
\intermediatestate{} state from being intermediately excited \emph{via
  the laser interaction}. The propagator \(\propUlaser\) for the laser
interaction \(\laserinteraction\) is replaced according to
\begin{equation}
  \label{eqn:projected-propagator}
  \propUlaser \to \proj\propUlaser\proj + \rej,
\end{equation}
where \(\rej\) is the projector onto the \intermediatestate{} state
and \(\proj\equiv\identity-\rej\) is the projector onto the orthogonal
complement. In this way, the \intermediatestate{} state is still
present in the calculation, but it will not be coupled via the laser
field; we can do this since in the length gauge, the field-free
excited state remains a good approximation to the time-dependent
eigenstate. The state chosen has \(\sim\num{0.979}\) contribution from
the \channela{} manifold, \(\sim\num{1.95e-2}\) contribution from
\channelb{}, and \(\sim\num{1.82e-3}\) contribution from \channelc{}
through configuration interaction, which makes it a likely candidate
for the antenna mechanism.

As we see in Figure~\ref{fig:coherence-rydberg-removed}, the degree of
coherence is strongly altered by the removal of \intermediatestate{},
confirming the importance of the Rydberg states in the formation of
the antenna. The exact influence of individual states on the antenna
efficiency and the final coherence is a topic for future
investigations.

\subsection{Antenna Transition Strength}
\begin{figure}[t]
  \centering
  \includegraphics{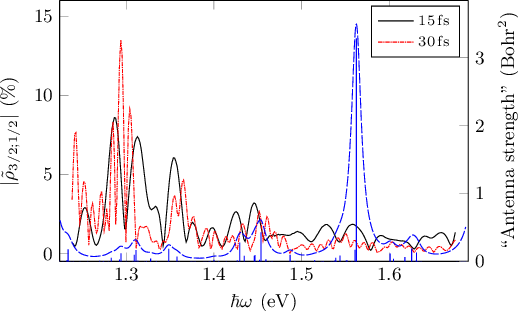}
  \caption{Antenna transitions: The left ordinate corresponds to the
    degree of coherence for \SI{15}{\femto\second} and
    \SI{30}{\femto\second} are shown in solid black and dot-dashed red
    lines, respectively. The right ordinate corresponds to the antenna
    strengths, computed using \eqref{eqn:antenna-strength} and shown
    as sticks, and convoluted with a Lorentzian spectral shape
    \eqref{eqn:lorentzian}, shown as the dashed blue line.}
  \label{fig:antenna-transitions}
\end{figure}

We now would like to investigate whether there is a correlation
between the transitions in the Rydberg manifold that constitute our
antenna, and the observed variation of the degree of coherence
\(\degreeofcoherence{3/2;1/2}\) with the photon energy. The weight of
the antenna transition between states \(a\) and \(b\) is estimated as
\begin{equation}
  \label{eqn:antenna-strength}
  w_{ab} =
  \abs{z_{ab}}^2[\min(\abs*{c^{(a)}_{3/2}}^2,\abs*{c^{(a)}_{1/2}}^2) +
  \min(\abs*{c^{(b)}_{3/2}}^2,\abs*{c^{(b)}_{1/2}}^2)],
\end{equation}
where \(c^{(s)}_{J}\) is the complex amplitude of state \(s\) in channel
\(J\). Diagonalizing the field-free Hamiltonian
[\eqref{eqn:td-cis-eoms-simplified} with \(\laserinteraction=0\)], we
obtain the first 150 excited states, and compute
\eqref{eqn:antenna-strength} for all dipole-allowed transitions. Those
that fall within the energy interval we consider, are shown as a stick
spectrum in Figure~\ref{fig:antenna-transitions}, alongside the degree
of coherence. By convoluting the stick spectrum with a Lorentzian
\begin{equation}
  \label{eqn:lorentzian}
  \Lambda(\omega) = \frac{1}{1+x^2}, \quad
  x = \frac{2\omega}{\Gamma},
\end{equation}
where \(\Gamma\) is the full width at half maximum, a continuous
distribution is acquired; we use \(\Gamma=\SI{5e-4}{\hartree}\),
\(\planck/\Gamma\sim\SI{48}{\femto\second}\).
The similarity of the convoluted spectrum with the degree of coherence
is very suggestive, apart from the very strong peak at
\(\sim\SI{1.56}{\electronvolt}\), which is due to a very strong dipole
moment for that transition. Exact agreement can, however, not be
expected for a variety of
reasons. Equation~\eqref{eqn:antenna-strength} considers dipole
transitions between field-free states, i.e.\ disregarding any Stark
shifts in the strong field, which means the transitions might not
occur at the positions indicated. More important, though, is the fact
that we completely disregard the relative populations of the
constituent states, which, when prepared through frustrated tunnelling
depend strongly on the laser parameters \cite{Nubbemeyer2008}.

\subsection{Antenna Size}
We now wish to estimate the effective size of the antenna structure,
and relate that to the driving wavelength. In classical
electromagnetic theory, a dipole antenna will exhibit the largest gain
if the length is \(5\lambda/4\); \(\lambda/2\) is also very common. Naturally,
electron excursions on that scale would far exceed the applicability
of the dipole approximation, however, this gives a clear motivation
for why large electronic structures are desirable to efficiently
couple the external electric field into the atom.

An excited state can in the CIS \emph{Ansatz} be written as
\begin{equation*}
  \sum_k\contslater{k},
\end{equation*}
with the particle orbital \(\contket*{k}\) containing all information
about the electron in the channel associated with
excitation/ionization from the occupied orbital \(\ket{k}\). We
estimate the size of the state as
\begin{equation*}
  s_1 \defd \sqrt{\sum_k\matrixel*{\cont{k}}{r^2}{\cont{k}}};
  \quad
  s_2 \defd \sum_k\sqrt{\matrixel*{\cont{k}}{r^2}{\cont{k}}}.
\end{equation*}
The size of the antenna is then estimated as the geometric mean of the
sizes of the two states:
\begin{equation*}
  \sqrt{s(a)s(b)}
\end{equation*}
For the transitions in Figure~\ref{fig:antenna-transitions}, the
estimates fall in the range \SIrange{1}{3}{\nano\meter}, and with a
driving wavelength of \(\lambda\sim\SI{900}{\nano\meter}\), this corresponds to
\(\frac{\lambda}{900}-\frac{\lambda}{300}\) antenna structures. This is of course
far from the optimum \(\frac{5\lambda}{4}\), but a lot better than what
could be expected from the orbitals of the ground state;
\(\conf{5\{s,p\}}\) have a size of \(\sim\SI{0.1}{\nano\meter}\) which
would yield a \(\frac{\lambda}{9000}\) antenna.

\subsection{Coherence due to Single Rydberg States}
\begin{figure}[tb]
  \centering
  \includegraphics{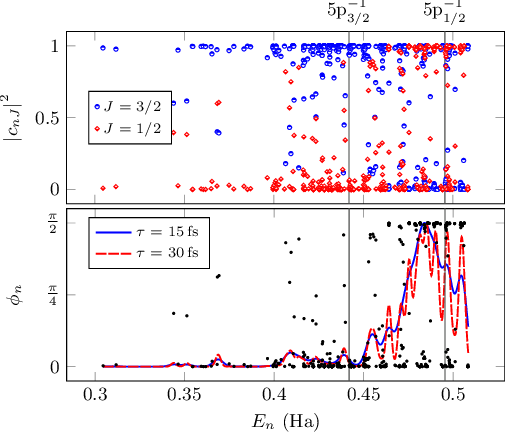}
  \caption{Rydberg states channel decomposition for the 500 first
    excited states of xenon; the top panel shows the populations of
    state \(n\) in \channela{} (blue circles) and \channelb{} (red
    diamonds), respectively, as a function of its excitation energy
    \(E_n\). \channelc{} contributions are negligible for these
    states. The vertical lines indicate the positions of the
    ionization thresholds. The bottom panel shows the mixing angle
    \(\phi_n\defd\arctan(\abs{c_{n;1/2}}/\abs{c_{n;3/2}})\);
    \(\phi_n=0\) indicates a state purely in \channela{},
    \(\phi_n=\pi/2\) indicates a state purely in \channelb{}, and
    \(\phi_n=\pi/4\) indicates an even mixture. The lines show the average
    mixing angle as a function of excitation energy, when convolving
    with a Gaussian corresponding to \SI{15}{\femto\second} duration
    (solid blue), and \SI{30}{\femto\second} duration (dashed red).}
  \label{fig:rydberg-states-channel-decomposition}
\end{figure}
Through resonant excitation, it is possible to generate high degrees
of coherence, since some Rydberg states have large mixing fractions in
\channela{} and \channelb{}. If an excited state has equal amplitudes
in the two channels, tracing out the excited electron would yield an
ionic superposition with \SI{100}{\percent} degree of coherence. By
choosing the excited state judiciously, we can thus achieve any
desired degree of coherence from \SIrange{0}{100}{\percent}. In
Figure~\ref{fig:rydberg-states-channel-decomposition}, we show the
mixing coefficients of the first 500 excited states of xenon. Below
the \channela{} threshold, the \(J=3/2\) component is dominant, with
only a few states achieving large fractions of \(J=1/2\). Between the
thresholds, the \(J=1/2\) component becomes more important. It is
precisely the latter states that \textcite{Dill1973} considered,
studying the importance of the spin--orbit interaction in
photoionization.

Can resonant excitation of an intermediate state with high mixing
between \channela{} and \channelb{} explain our observed degree of
coherence in Figure~\ref{fig:degree-of-coherence}? Let us first
consider weak-field ionization, where we first through one-photon
absorption populate the intermediate state with energy \(E_n\), which
we may write as
\begin{equation}
  \label{eqn:intermediate-state-weak-field}
  \ket{\Psi_n} = \antisym(c_I\ket{I}\ket{n_I} + c_J\ket{J}\ket{n_J}),
\end{equation}
where \(\antisym\) is the antisymmetrization operator. Subsequent
single-photon ionization will lead to a final state on the form
\eqref{eqn:ion-electron-state}. However, even if \(c_I\) and \(c_J\)
in \eqref{eqn:intermediate-state-weak-field} are both significant,
\(\eloverlap=\braket*{\chi_I}{\chi_J}\) in \eqref{eqn:ion-electron-state}
will still vanish due to energy conservation; the photoelectron peaks
will appear at \(W_k=\omega-(\ionpotential[I]-E_n)\) and
\(W_k=\omega-(\ionpotential[J]-E_n)\), respectively. This is not the case
in the process considered by \textcite{Dill1973}, since the final
state involves only one ion channel, namely \channela{}, which is
populated through direct ionization, as well as autoionization of the
intermediately excited states below the \channelb{} threshold. Thus
energy conservation is automatically fulfilled.

We next consider strong-field ionization. In this case, it is
difficult to address a single state. Instead, we access the average
coherence of the state manifold, which remains low; see the average
mixing angle in the lower panel of
Figure~\ref{fig:rydberg-states-channel-decomposition}. Furthermore,
subsequent ionization and generation of ATI progressions would still
face the same predicament as stated earlier: for \(\eta\sim1\), the
photoelectron peaks of similar kinetic energy would result from
absorption of a different number of photons, and thus by parity, their
overlap would vanish. The atomic antenna, which repeatedly accesses
parts of the excited spectrum with high mixing fractions, allows us to
amplify this small, average mixing coefficient.

\vfill

\bibliography{exported-bibliography}

\begin{thebibliography}{47}%
\makeatletter
\providecommand \@ifxundefined [1]{%
 \@ifx{#1\undefined}
}%
\providecommand \@ifnum [1]{%
 \ifnum #1\expandafter \@firstoftwo
 \else \expandafter \@secondoftwo
 \fi
}%
\providecommand \@ifx [1]{%
 \ifx #1\expandafter \@firstoftwo
 \else \expandafter \@secondoftwo
 \fi
}%
\providecommand \natexlab [1]{#1}%
\providecommand \enquote  [1]{``#1''}%
\providecommand \bibnamefont  [1]{#1}%
\providecommand \bibfnamefont [1]{#1}%
\providecommand \citenamefont [1]{#1}%
\providecommand \href@noop [0]{\@secondoftwo}%
\providecommand \href [0]{\begingroup \@sanitize@url \@href}%
\providecommand \@href[1]{\@@startlink{#1}\@@href}%
\providecommand \@@href[1]{\endgroup#1\@@endlink}%
\providecommand \@sanitize@url [0]{\catcode `\\12\catcode `\$12\catcode
  `\&12\catcode `\#12\catcode `\^12\catcode `\_12\catcode `\%12\relax}%
\providecommand \@@startlink[1]{}%
\providecommand \@@endlink[0]{}%
\providecommand \url  [0]{\begingroup\@sanitize@url \@url }%
\providecommand \@url [1]{\endgroup\@href {#1}{\urlprefix }}%
\providecommand \urlprefix  [0]{URL }%
\providecommand \Eprint [0]{\href }%
\providecommand \doibase [0]{https://doi.org/}%
\providecommand \selectlanguage [0]{\@gobble}%
\providecommand \bibinfo  [0]{\@secondoftwo}%
\providecommand \bibfield  [0]{\@secondoftwo}%
\providecommand \translation [1]{[#1]}%
\providecommand \BibitemOpen [0]{}%
\providecommand \bibitemStop [0]{}%
\providecommand \bibitemNoStop [0]{.\EOS\space}%
\providecommand \EOS [0]{\spacefactor3000\relax}%
\providecommand \BibitemShut  [1]{\csname bibitem#1\endcsname}%
\let\auto@bib@innerbib\@empty
\bibitem [{\citenamefont {Wragg}\ \emph {et~al.}(2019)\citenamefont {Wragg},
  \citenamefont {Clarke}, \citenamefont {Armstrong}, \citenamefont {Brown},
  \citenamefont {Ballance},\ and\ \citenamefont {van~der Hart}}]{Wragg2019}%
  \BibitemOpen
  \bibfield  {author} {\bibinfo {author} {\bibfnamefont {J.}~\bibnamefont
  {Wragg}}, \bibinfo {author} {\bibfnamefont {D.~D.~A.}\ \bibnamefont
  {Clarke}}, \bibinfo {author} {\bibfnamefont {G.~S.~J.}\ \bibnamefont
  {Armstrong}}, \bibinfo {author} {\bibfnamefont {A.~C.}\ \bibnamefont
  {Brown}}, \bibinfo {author} {\bibfnamefont {C.~P.}\ \bibnamefont
  {Ballance}},\ and\ \bibinfo {author} {\bibfnamefont {H.~W.}\ \bibnamefont
  {van~der Hart}},\ }\bibfield  {title} {\bibinfo {title} {Resolving ultrafast
  spin--orbit dynamics in heavy many-electron atoms},\ }\href
  {https://doi.org/10.1103/physrevlett.123.163001} {\bibfield  {journal}
  {\bibinfo  {journal} {Physical Review Letters}\ }\textbf {\bibinfo {volume}
  {123}},\ \bibinfo {pages} {163001} (\bibinfo {year} {2019})}\BibitemShut
  {NoStop}%
\bibitem [{\citenamefont {K{\"u}bel}\ \emph {et~al.}(2019)\citenamefont
  {K{\"u}bel}, \citenamefont {Dube}, \citenamefont {Naumov}, \citenamefont
  {Villeneuve}, \citenamefont {Corkum},\ and\ \citenamefont
  {Staudte}}]{Kuebel2019}%
  \BibitemOpen
  \bibfield  {author} {\bibinfo {author} {\bibfnamefont {M.}~\bibnamefont
  {K{\"u}bel}}, \bibinfo {author} {\bibfnamefont {Z.}~\bibnamefont {Dube}},
  \bibinfo {author} {\bibfnamefont {A.~Y.}\ \bibnamefont {Naumov}}, \bibinfo
  {author} {\bibfnamefont {D.~M.}\ \bibnamefont {Villeneuve}}, \bibinfo
  {author} {\bibfnamefont {P.~B.}\ \bibnamefont {Corkum}},\ and\ \bibinfo
  {author} {\bibfnamefont {A.}~\bibnamefont {Staudte}},\ }\bibfield  {title}
  {\bibinfo {title} {Spatiotemporal imaging of valence electron motion},\
  }\href {https://doi.org/10.1038/s41467-019-09036-w} {\bibfield  {journal}
  {\bibinfo  {journal} {Nature Communications}\ }\textbf {\bibinfo {volume}
  {10}},\ \bibinfo {pages} {1042} (\bibinfo {year} {2019})}\BibitemShut
  {NoStop}%
\bibitem [{\citenamefont {Goulielmakis}\ \emph {et~al.}(2010)\citenamefont
  {Goulielmakis}, \citenamefont {Loh}, \citenamefont {Wirth}, \citenamefont
  {Santra}, \citenamefont {Rohringer}, \citenamefont {Yakovlev}, \citenamefont
  {Zherebtsov}, \citenamefont {Pfeifer}, \citenamefont {Azzeer}, \citenamefont
  {Kling}, \citenamefont {Leone},\ and\ \citenamefont
  {Krausz}}]{Goulielmakis2010}%
  \BibitemOpen
  \bibfield  {author} {\bibinfo {author} {\bibfnamefont {E.}~\bibnamefont
  {Goulielmakis}}, \bibinfo {author} {\bibfnamefont {Z.-H.}\ \bibnamefont
  {Loh}}, \bibinfo {author} {\bibfnamefont {A.}~\bibnamefont {Wirth}}, \bibinfo
  {author} {\bibfnamefont {R.}~\bibnamefont {Santra}}, \bibinfo {author}
  {\bibfnamefont {N.}~\bibnamefont {Rohringer}}, \bibinfo {author}
  {\bibfnamefont {V.~S.}\ \bibnamefont {Yakovlev}}, \bibinfo {author}
  {\bibfnamefont {S.}~\bibnamefont {Zherebtsov}}, \bibinfo {author}
  {\bibfnamefont {T.}~\bibnamefont {Pfeifer}}, \bibinfo {author} {\bibfnamefont
  {A.~M.}\ \bibnamefont {Azzeer}}, \bibinfo {author} {\bibfnamefont {M.~F.}\
  \bibnamefont {Kling}}, \bibinfo {author} {\bibfnamefont {S.~R.}\ \bibnamefont
  {Leone}},\ and\ \bibinfo {author} {\bibfnamefont {F.}~\bibnamefont
  {Krausz}},\ }\bibfield  {title} {\bibinfo {title} {Real-time observation of
  valence electron motion},\ }\href {https://doi.org/10.1038/nature09212}
  {\bibfield  {journal} {\bibinfo  {journal} {Nature}\ }\textbf {\bibinfo
  {volume} {466}},\ \bibinfo {pages} {739} (\bibinfo {year}
  {2010})}\BibitemShut {NoStop}%
\bibitem [{\citenamefont {Hartung}\ \emph {et~al.}(2016)\citenamefont
  {Hartung}, \citenamefont {Morales}, \citenamefont {Kunitski}, \citenamefont
  {Henrichs}, \citenamefont {Laucke}, \citenamefont {Richter}, \citenamefont
  {Jahnke}, \citenamefont {Kalinin}, \citenamefont {Schöffler}, \citenamefont
  {Schmidt},\ and\ \citenamefont {et~al.}}]{Hartung2016}%
  \BibitemOpen
  \bibfield  {author} {\bibinfo {author} {\bibfnamefont {A.}~\bibnamefont
  {Hartung}}, \bibinfo {author} {\bibfnamefont {F.}~\bibnamefont {Morales}},
  \bibinfo {author} {\bibfnamefont {M.}~\bibnamefont {Kunitski}}, \bibinfo
  {author} {\bibfnamefont {K.}~\bibnamefont {Henrichs}}, \bibinfo {author}
  {\bibfnamefont {A.}~\bibnamefont {Laucke}}, \bibinfo {author} {\bibfnamefont
  {M.}~\bibnamefont {Richter}}, \bibinfo {author} {\bibfnamefont
  {T.}~\bibnamefont {Jahnke}}, \bibinfo {author} {\bibfnamefont
  {A.}~\bibnamefont {Kalinin}}, \bibinfo {author} {\bibfnamefont
  {M.}~\bibnamefont {Schöffler}}, \bibinfo {author} {\bibfnamefont {L.~P.~H.}\
  \bibnamefont {Schmidt}},\ and\ \bibinfo {author} {\bibnamefont {et~al.}},\
  }\bibfield  {title} {\bibinfo {title} {Electron spin polarization in
  strong-field ionization of xenon atoms},\ }\href
  {https://doi.org/10.1038/nphoton.2016.109} {\bibfield  {journal} {\bibinfo
  {journal} {Nature Photonics}\ }\textbf {\bibinfo {volume} {10}},\ \bibinfo
  {pages} {526} (\bibinfo {year} {2016})}\BibitemShut {NoStop}%
\bibitem [{\citenamefont {Barth}\ and\ \citenamefont
  {Smirnova}(2013)}]{Barth2013}%
  \BibitemOpen
  \bibfield  {author} {\bibinfo {author} {\bibfnamefont {I.}~\bibnamefont
  {Barth}}\ and\ \bibinfo {author} {\bibfnamefont {O.}~\bibnamefont
  {Smirnova}},\ }\bibfield  {title} {\bibinfo {title} {Spin-polarized electrons
  produced by strong-field ionization},\ }\href
  {https://doi.org/10.1103/physreva.88.013401} {\bibfield  {journal} {\bibinfo
  {journal} {Physical Review A}\ }\textbf {\bibinfo {volume} {88}},\ \bibinfo
  {pages} {013401} (\bibinfo {year} {2013})}\BibitemShut {NoStop}%
\bibitem [{\citenamefont {Barth}\ and\ \citenamefont
  {Smirnova}(2014)}]{Barth2014}%
  \BibitemOpen
  \bibfield  {author} {\bibinfo {author} {\bibfnamefont {I.}~\bibnamefont
  {Barth}}\ and\ \bibinfo {author} {\bibfnamefont {O.}~\bibnamefont
  {Smirnova}},\ }\bibfield  {title} {\bibinfo {title} {Hole dynamics and spin
  currents after ionization in strong circularly polarized laser fields},\
  }\href {https://doi.org/10.1088/0953-4075/47/20/204020} {\bibfield  {journal}
  {\bibinfo  {journal} {Journal of Physics B: Atomic, Molecular and Optical
  Physics}\ }\textbf {\bibinfo {volume} {47}},\ \bibinfo {pages} {204020}
  (\bibinfo {year} {2014})}\BibitemShut {NoStop}%
\bibitem [{\citenamefont {Kuchiev}(1987)}]{Kuchiev1987}%
  \BibitemOpen
  \bibfield  {author} {\bibinfo {author} {\bibfnamefont {M.~Y.}\ \bibnamefont
  {Kuchiev}},\ }\bibfield  {title} {\bibinfo {title} {Atomic antenna},\ }\href
  {http://jetpletters.ru/ps/1241/article_18763.shtml} {\bibfield  {journal}
  {\bibinfo  {journal} {Soviet Journal of Experimental and Theoretical Physics
  Letters}\ }\textbf {\bibinfo {volume} {45}},\ \bibinfo {pages} {404}
  (\bibinfo {year} {1987})}\BibitemShut {NoStop}%
\bibitem [{\citenamefont {Tzur}\ \emph {et~al.}(2021)\citenamefont {Tzur},
  \citenamefont {Neufeld}, \citenamefont {Fleischer},\ and\ \citenamefont
  {Cohen}}]{Tzur2021}%
  \BibitemOpen
  \bibfield  {author} {\bibinfo {author} {\bibfnamefont {M.~E.}\ \bibnamefont
  {Tzur}}, \bibinfo {author} {\bibfnamefont {O.}~\bibnamefont {Neufeld}},
  \bibinfo {author} {\bibfnamefont {A.}~\bibnamefont {Fleischer}},\ and\
  \bibinfo {author} {\bibfnamefont {O.}~\bibnamefont {Cohen}},\ }\bibfield
  {title} {\bibinfo {title} {Selection rules for breaking selection rules},\
  }\href {https://doi.org/10.1088/1367-2630/ac27e2} {\bibfield  {journal}
  {\bibinfo  {journal} {New Journal of Physics}\ }\textbf {\bibinfo {volume}
  {23}},\ \bibinfo {pages} {103039} (\bibinfo {year} {2021})}\BibitemShut
  {NoStop}%
\bibitem [{\citenamefont {Nesbet}(1955)}]{Nesbet1955}%
  \BibitemOpen
  \bibfield  {author} {\bibinfo {author} {\bibfnamefont {R.~K.}\ \bibnamefont
  {Nesbet}},\ }\bibfield  {title} {\bibinfo {title} {Configuration interaction
  in orbital theories},\ }\href {https://doi.org/10.1098/rspa.1955.0134}
  {\bibfield  {journal} {\bibinfo  {journal} {Proceedings of the Royal Society
  of London. Series A. Mathematical and Physical Sciences}\ }\textbf {\bibinfo
  {volume} {230}},\ \bibinfo {pages} {312} (\bibinfo {year}
  {1955})}\BibitemShut {NoStop}%
\bibitem [{\citenamefont {L{\"o}wdin}(1955)}]{Loewdin1955a}%
  \BibitemOpen
  \bibfield  {author} {\bibinfo {author} {\bibfnamefont {P.-O.}\ \bibnamefont
  {L{\"o}wdin}},\ }\bibfield  {title} {\bibinfo {title} {Quantum theory of
  many-particle systems. {I}. {P}hysical interpretations by means of density
  matrices, natural spin-orbitals, and convergence problems in the method of
  configurational interaction},\ }\href
  {https://doi.org/10.1103/physrev.97.1474} {\bibfield  {journal} {\bibinfo
  {journal} {Physical Review}\ }\textbf {\bibinfo {volume} {97}},\ \bibinfo
  {pages} {1474} (\bibinfo {year} {1955})}\BibitemShut {NoStop}%
\bibitem [{\citenamefont {Krause}\ \emph {et~al.}(2005)\citenamefont {Krause},
  \citenamefont {Klamroth},\ and\ \citenamefont {Saalfrank}}]{Krause2005}%
  \BibitemOpen
  \bibfield  {author} {\bibinfo {author} {\bibfnamefont {P.}~\bibnamefont
  {Krause}}, \bibinfo {author} {\bibfnamefont {T.}~\bibnamefont {Klamroth}},\
  and\ \bibinfo {author} {\bibfnamefont {P.}~\bibnamefont {Saalfrank}},\
  }\bibfield  {title} {\bibinfo {title} {Time-dependent
  configuration-interaction calculations of laser-pulse-driven many-electron
  dynamics: Controlled dipole switching in lithium cyanide},\ }\href
  {https://doi.org/10.1063/1.1999636} {\bibfield  {journal} {\bibinfo
  {journal} {The Journal of Chemical Physics}\ }\textbf {\bibinfo {volume}
  {123}},\ \bibinfo {pages} {074105} (\bibinfo {year} {2005})}\BibitemShut
  {NoStop}%
\bibitem [{\citenamefont {Rohringer}\ \emph {et~al.}(2006)\citenamefont
  {Rohringer}, \citenamefont {Gordon},\ and\ \citenamefont
  {Santra}}]{Rohringer2006}%
  \BibitemOpen
  \bibfield  {author} {\bibinfo {author} {\bibfnamefont {N.}~\bibnamefont
  {Rohringer}}, \bibinfo {author} {\bibfnamefont {A.}~\bibnamefont {Gordon}},\
  and\ \bibinfo {author} {\bibfnamefont {R.}~\bibnamefont {Santra}},\
  }\bibfield  {title} {\bibinfo {title} {Configuration-interaction-based
  time-dependent orbital approach for \emph{Ab Initio} treatment of electronic
  dynamics in a strong optical laser field},\ }\href
  {https://doi.org/10.1103/physreva.74.043420} {\bibfield  {journal} {\bibinfo
  {journal} {Physical Review A}\ }\textbf {\bibinfo {volume} {74}},\ \bibinfo
  {pages} {043420} (\bibinfo {year} {2006})}\BibitemShut {NoStop}%
\bibitem [{\citenamefont {Greenman}\ \emph {et~al.}(2010)\citenamefont
  {Greenman}, \citenamefont {Ho}, \citenamefont {Pabst}, \citenamefont
  {Kamarchik}, \citenamefont {Mazziotti},\ and\ \citenamefont
  {Santra}}]{Greenman2010PRA}%
  \BibitemOpen
  \bibfield  {author} {\bibinfo {author} {\bibfnamefont {L.}~\bibnamefont
  {Greenman}}, \bibinfo {author} {\bibfnamefont {P.~J.}\ \bibnamefont {Ho}},
  \bibinfo {author} {\bibfnamefont {S.}~\bibnamefont {Pabst}}, \bibinfo
  {author} {\bibfnamefont {E.}~\bibnamefont {Kamarchik}}, \bibinfo {author}
  {\bibfnamefont {D.~A.}\ \bibnamefont {Mazziotti}},\ and\ \bibinfo {author}
  {\bibfnamefont {R.}~\bibnamefont {Santra}},\ }\bibfield  {title} {\bibinfo
  {title} {Implementation of the time-dependent configuration-interaction
  singles method for atomic strong-field processes},\ }\href
  {https://doi.org/10.1103/physreva.82.023406} {\bibfield  {journal} {\bibinfo
  {journal} {Physical Review A}\ }\textbf {\bibinfo {volume} {82}},\ \bibinfo
  {pages} {023406} (\bibinfo {year} {2010})}\BibitemShut {NoStop}%
\bibitem [{\citenamefont {Carlström}\ \emph
  {et~al.}(2022{\natexlab{a}})\citenamefont {Carlström}, \citenamefont
  {Spanner},\ and\ \citenamefont {Patchkovskii}}]{Carlstroem2022tdcisI}%
  \BibitemOpen
  \bibfield  {author} {\bibinfo {author} {\bibfnamefont {S.}~\bibnamefont
  {Carlström}}, \bibinfo {author} {\bibfnamefont {M.}~\bibnamefont
  {Spanner}},\ and\ \bibinfo {author} {\bibfnamefont {S.}~\bibnamefont
  {Patchkovskii}},\ }\bibfield  {title} {\bibinfo {title} {General
  time-dependent configuration-interaction singles {I}: The molecular case},\
  }\Eprint {https://arxiv.org/abs/2204.09966} {arXiv:2204.09966
  [physics.chem-ph]}  (\bibinfo {year} {2022}{\natexlab{a}}),\ \bibinfo {note}
  {accepted for publication in \emph{Physical Review A}}\BibitemShut {NoStop}%
\bibitem [{\citenamefont {Carlström}\ \emph
  {et~al.}(2022{\natexlab{b}})\citenamefont {Carlström}, \citenamefont
  {Bertolino}, \citenamefont {Dahlström},\ and\ \citenamefont
  {Patchkovskii}}]{Carlstroem2022tdcisII}%
  \BibitemOpen
  \bibfield  {author} {\bibinfo {author} {\bibfnamefont {S.}~\bibnamefont
  {Carlström}}, \bibinfo {author} {\bibfnamefont {M.}~\bibnamefont
  {Bertolino}}, \bibinfo {author} {\bibfnamefont {J.~M.}\ \bibnamefont
  {Dahlström}},\ and\ \bibinfo {author} {\bibfnamefont {S.}~\bibnamefont
  {Patchkovskii}},\ }\bibfield  {title} {\bibinfo {title} {General
  time-dependent configuration-interaction singles {II}: The atomic case},\
  }\Eprint {https://arxiv.org/abs/2204.10534} {arXiv:2204.10534
  [physics.atom-ph]}  (\bibinfo {year} {2022}{\natexlab{b}}),\ \bibinfo {note}
  {accepted for publication in \emph{Physical Review A}}\BibitemShut {NoStop}%
\bibitem [{\citenamefont {Peterson}\ \emph {et~al.}(2003)\citenamefont
  {Peterson}, \citenamefont {Figgen}, \citenamefont {Goll}, \citenamefont
  {Stoll},\ and\ \citenamefont {Dolg}}]{Peterson2003}%
  \BibitemOpen
  \bibfield  {author} {\bibinfo {author} {\bibfnamefont {K.~A.}\ \bibnamefont
  {Peterson}}, \bibinfo {author} {\bibfnamefont {D.}~\bibnamefont {Figgen}},
  \bibinfo {author} {\bibfnamefont {E.}~\bibnamefont {Goll}}, \bibinfo {author}
  {\bibfnamefont {H.}~\bibnamefont {Stoll}},\ and\ \bibinfo {author}
  {\bibfnamefont {M.}~\bibnamefont {Dolg}},\ }\bibfield  {title} {\bibinfo
  {title} {Systematically convergent basis sets with relativistic
  pseudopotentials. {II}. {S}mall-core pseudopotentials and correlation
  consistent basis sets for the post-d group 16--18 elements},\ }\href
  {https://doi.org/10.1063/1.1622924} {\bibfield  {journal} {\bibinfo
  {journal} {The Journal of Chemical Physics}\ }\textbf {\bibinfo {volume}
  {119}},\ \bibinfo {pages} {11113} (\bibinfo {year} {2003})}\BibitemShut
  {NoStop}%
\bibitem [{\citenamefont {Zapata}\ \emph {et~al.}(2022)\citenamefont {Zapata},
  \citenamefont {Vinbladh}, \citenamefont {Ljungdahl}, \citenamefont
  {Lindroth},\ and\ \citenamefont {Dahlstr{\"o}m}}]{Zapata2022}%
  \BibitemOpen
  \bibfield  {author} {\bibinfo {author} {\bibfnamefont {F.}~\bibnamefont
  {Zapata}}, \bibinfo {author} {\bibfnamefont {J.}~\bibnamefont {Vinbladh}},
  \bibinfo {author} {\bibfnamefont {A.}~\bibnamefont {Ljungdahl}}, \bibinfo
  {author} {\bibfnamefont {E.}~\bibnamefont {Lindroth}},\ and\ \bibinfo
  {author} {\bibfnamefont {J.~M.}\ \bibnamefont {Dahlstr{\"o}m}},\ }\bibfield
  {title} {\bibinfo {title} {Relativistic time-dependent
  configuration-interaction singles method},\ }\href
  {https://doi.org/10.1103/physreva.105.012802} {\bibfield  {journal} {\bibinfo
   {journal} {Physical Review A}\ }\textbf {\bibinfo {volume} {105}},\ \bibinfo
  {pages} {012802} (\bibinfo {year} {2022})}\BibitemShut {NoStop}%
\bibitem [{\citenamefont {Ermolaev}\ \emph {et~al.}(1999)\citenamefont
  {Ermolaev}, \citenamefont {Puzynin}, \citenamefont {Selin},\ and\
  \citenamefont {Vinitsky}}]{Ermolaev1999}%
  \BibitemOpen
  \bibfield  {author} {\bibinfo {author} {\bibfnamefont {A.~M.}\ \bibnamefont
  {Ermolaev}}, \bibinfo {author} {\bibfnamefont {I.~V.}\ \bibnamefont
  {Puzynin}}, \bibinfo {author} {\bibfnamefont {A.~V.}\ \bibnamefont {Selin}},\
  and\ \bibinfo {author} {\bibfnamefont {S.~I.}\ \bibnamefont {Vinitsky}},\
  }\bibfield  {title} {\bibinfo {title} {Integral boundary conditions for the
  time-dependent {S}chr{\"o}dinger equation: Atom in a laser field},\ }\href
  {https://doi.org/10.1103/physreva.60.4831} {\bibfield  {journal} {\bibinfo
  {journal} {Physical Review A}\ }\textbf {\bibinfo {volume} {60}},\ \bibinfo
  {pages} {4831} (\bibinfo {year} {1999})}\BibitemShut {NoStop}%
\bibitem [{\citenamefont {Ermolaev}\ and\ \citenamefont
  {Selin}(2000)}]{Ermolaev2000}%
  \BibitemOpen
  \bibfield  {author} {\bibinfo {author} {\bibfnamefont {A.~M.}\ \bibnamefont
  {Ermolaev}}\ and\ \bibinfo {author} {\bibfnamefont {A.~V.}\ \bibnamefont
  {Selin}},\ }\bibfield  {title} {\bibinfo {title} {Integral boundary
  conditions for the time-dependent {S}chr{\"o}dinger equation: Superposition
  of the laser field and a long-range atomic potential},\ }\href
  {https://doi.org/10.1103/physreva.62.015401} {\bibfield  {journal} {\bibinfo
  {journal} {Physical Review A}\ }\textbf {\bibinfo {volume} {62}},\ \bibinfo
  {pages} {015401} (\bibinfo {year} {2000})}\BibitemShut {NoStop}%
\bibitem [{\citenamefont {Serov}\ \emph {et~al.}(2001)\citenamefont {Serov},
  \citenamefont {Derbov}, \citenamefont {Joulakian},\ and\ \citenamefont
  {Vinitsky}}]{Serov2001}%
  \BibitemOpen
  \bibfield  {author} {\bibinfo {author} {\bibfnamefont {V.~V.}\ \bibnamefont
  {Serov}}, \bibinfo {author} {\bibfnamefont {V.~L.}\ \bibnamefont {Derbov}},
  \bibinfo {author} {\bibfnamefont {B.~B.}\ \bibnamefont {Joulakian}},\ and\
  \bibinfo {author} {\bibfnamefont {S.~I.}\ \bibnamefont {Vinitsky}},\
  }\bibfield  {title} {\bibinfo {title} {Wave-packet evolution approach to
  ionization of the hydrogen molecular ion by fast electrons},\ }\href
  {https://doi.org/10.1103/physreva.63.062711} {\bibfield  {journal} {\bibinfo
  {journal} {Physical Review A}\ }\textbf {\bibinfo {volume} {63}},\ \bibinfo
  {pages} {062711} (\bibinfo {year} {2001})}\BibitemShut {NoStop}%
\bibitem [{\citenamefont {Tao}\ and\ \citenamefont
  {Scrinzi}(2012)}]{Tao2012NJoP}%
  \BibitemOpen
  \bibfield  {author} {\bibinfo {author} {\bibfnamefont {L.}~\bibnamefont
  {Tao}}\ and\ \bibinfo {author} {\bibfnamefont {A.}~\bibnamefont {Scrinzi}},\
  }\bibfield  {title} {\bibinfo {title} {Photo-electron momentum spectra from
  minimal volumes: the time-dependent surface flux method},\ }\href
  {https://doi.org/10.1088/1367-2630/14/1/013021} {\bibfield  {journal}
  {\bibinfo  {journal} {New Journal of Physics}\ }\textbf {\bibinfo {volume}
  {14}},\ \bibinfo {pages} {013021} (\bibinfo {year} {2012})}\BibitemShut
  {NoStop}%
\bibitem [{\citenamefont {Scrinzi}(2012)}]{Scrinzi2012NJoP}%
  \BibitemOpen
  \bibfield  {author} {\bibinfo {author} {\bibfnamefont {A.}~\bibnamefont
  {Scrinzi}},\ }\bibfield  {title} {\bibinfo {title} {t-{SURFF}: {F}ully
  {D}ifferential {T}wo-{E}lectron {P}hoto-{E}mission {S}pectra},\ }\href
  {https://doi.org/10.1088/1367-2630/14/8/085008} {\bibfield  {journal}
  {\bibinfo  {journal} {New Journal of Physics}\ }\textbf {\bibinfo {volume}
  {14}},\ \bibinfo {pages} {085008} (\bibinfo {year} {2012})}\BibitemShut
  {NoStop}%
\bibitem [{\citenamefont {Morales}\ \emph {et~al.}(2016)\citenamefont
  {Morales}, \citenamefont {Bredtmann},\ and\ \citenamefont
  {Patchkovskii}}]{Morales2016-isurf}%
  \BibitemOpen
  \bibfield  {author} {\bibinfo {author} {\bibfnamefont {F.}~\bibnamefont
  {Morales}}, \bibinfo {author} {\bibfnamefont {T.}~\bibnamefont {Bredtmann}},\
  and\ \bibinfo {author} {\bibfnamefont {S.}~\bibnamefont {Patchkovskii}},\
  }\bibfield  {title} {\bibinfo {title} {{iSURF}: a family of infinite-time
  surface flux methods},\ }\href
  {https://doi.org/10.1088/0953-4075/49/24/245001} {\bibfield  {journal}
  {\bibinfo  {journal} {Journal of Physics B: Atomic, Molecular and Optical
  Physics}\ }\textbf {\bibinfo {volume} {49}},\ \bibinfo {pages} {245001}
  (\bibinfo {year} {2016})}\BibitemShut {NoStop}%
\bibitem [{\citenamefont {Pabst}\ \emph {et~al.}(2016)\citenamefont {Pabst},
  \citenamefont {Lein},\ and\ \citenamefont {Wörner}}]{Pabst2016}%
  \BibitemOpen
  \bibfield  {author} {\bibinfo {author} {\bibfnamefont {S.}~\bibnamefont
  {Pabst}}, \bibinfo {author} {\bibfnamefont {M.}~\bibnamefont {Lein}},\ and\
  \bibinfo {author} {\bibfnamefont {H.~J.}\ \bibnamefont {Wörner}},\
  }\bibfield  {title} {\bibinfo {title} {Preparing attosecond coherences by
  strong-field ionization},\ }\href
  {https://doi.org/10.1103/physreva.93.023412} {\bibfield  {journal} {\bibinfo
  {journal} {Physical Review A}\ }\textbf {\bibinfo {volume} {93}},\ \bibinfo
  {pages} {023412} (\bibinfo {year} {2016})}\BibitemShut {NoStop}%
\bibitem [{\citenamefont {Ruberti}\ \emph {et~al.}(2018)\citenamefont
  {Ruberti}, \citenamefont {Decleva},\ and\ \citenamefont
  {Averbukh}}]{Ruberti2018}%
  \BibitemOpen
  \bibfield  {author} {\bibinfo {author} {\bibfnamefont {M.}~\bibnamefont
  {Ruberti}}, \bibinfo {author} {\bibfnamefont {P.}~\bibnamefont {Decleva}},\
  and\ \bibinfo {author} {\bibfnamefont {V.}~\bibnamefont {Averbukh}},\
  }\bibfield  {title} {\bibinfo {title} {Full \emph{Ab Initio} many-electron
  simulation of attosecond molecular pump-probe spectroscopy},\ }\href
  {https://doi.org/10.1021/acs.jctc.8b00479} {\bibfield  {journal} {\bibinfo
  {journal} {Journal of Chemical Theory and Computation}\ }\textbf {\bibinfo
  {volume} {14}},\ \bibinfo {pages} {4991} (\bibinfo {year}
  {2018})}\BibitemShut {NoStop}%
\bibitem [{\citenamefont {Ruberti}(2019)}]{Ruberti2019a}%
  \BibitemOpen
  \bibfield  {author} {\bibinfo {author} {\bibfnamefont {M.}~\bibnamefont
  {Ruberti}},\ }\bibfield  {title} {\bibinfo {title} {Onset of ionic coherence
  and ultrafast charge dynamics in attosecond molecular ionisation},\ }\href
  {https://doi.org/10.1039/c9cp03074c} {\bibfield  {journal} {\bibinfo
  {journal} {Physical Chemistry Chemical Physics}\ }\textbf {\bibinfo {volume}
  {21}},\ \bibinfo {pages} {17584} (\bibinfo {year} {2019})}\BibitemShut
  {NoStop}%
\bibitem [{\citenamefont {Ruberti}(2021)}]{Ruberti2021}%
  \BibitemOpen
  \bibfield  {author} {\bibinfo {author} {\bibfnamefont {M.}~\bibnamefont
  {Ruberti}},\ }\bibfield  {title} {\bibinfo {title} {Quantum electronic
  coherences by attosecond transient absorption spectroscopy: \emph{Ab Initio}
  {B}-spline {RCS-ADC} study},\ }\href {https://doi.org/10.1039/d0fd00104j}
  {\bibfield  {journal} {\bibinfo  {journal} {Faraday Discussions}\ }\textbf
  {\bibinfo {volume} {228}},\ \bibinfo {pages} {286} (\bibinfo {year}
  {2021})}\BibitemShut {NoStop}%
\bibitem [{\citenamefont {Nubbemeyer}\ \emph {et~al.}(2008)\citenamefont
  {Nubbemeyer}, \citenamefont {Gorling}, \citenamefont {Saenz}, \citenamefont
  {Eichmann},\ and\ \citenamefont {Sandner}}]{Nubbemeyer2008}%
  \BibitemOpen
  \bibfield  {author} {\bibinfo {author} {\bibfnamefont {T.}~\bibnamefont
  {Nubbemeyer}}, \bibinfo {author} {\bibfnamefont {K.}~\bibnamefont {Gorling}},
  \bibinfo {author} {\bibfnamefont {A.}~\bibnamefont {Saenz}}, \bibinfo
  {author} {\bibfnamefont {U.}~\bibnamefont {Eichmann}},\ and\ \bibinfo
  {author} {\bibfnamefont {W.}~\bibnamefont {Sandner}},\ }\bibfield  {title}
  {\bibinfo {title} {Strong-field tunneling without ionization},\ }\href
  {https://doi.org/10.1103/physrevlett.101.233001} {\bibfield  {journal}
  {\bibinfo  {journal} {Physical Review Letters}\ }\textbf {\bibinfo {volume}
  {101}},\ \bibinfo {pages} {233001} (\bibinfo {year} {2008})}\BibitemShut
  {NoStop}%
\bibitem [{\citenamefont {Eichmann}\ \emph {et~al.}(2009)\citenamefont
  {Eichmann}, \citenamefont {Nubbemeyer}, \citenamefont {Rottke},\ and\
  \citenamefont {Sandner}}]{Eichmann2009}%
  \BibitemOpen
  \bibfield  {author} {\bibinfo {author} {\bibfnamefont {U.}~\bibnamefont
  {Eichmann}}, \bibinfo {author} {\bibfnamefont {T.}~\bibnamefont
  {Nubbemeyer}}, \bibinfo {author} {\bibfnamefont {H.}~\bibnamefont {Rottke}},\
  and\ \bibinfo {author} {\bibfnamefont {W.}~\bibnamefont {Sandner}},\
  }\bibfield  {title} {\bibinfo {title} {Acceleration of neutral atoms in
  strong short-pulse laser fields},\ }\href
  {https://doi.org/10.1038/nature08481} {\bibfield  {journal} {\bibinfo
  {journal} {Nature}\ }\textbf {\bibinfo {volume} {461}},\ \bibinfo {pages}
  {1261} (\bibinfo {year} {2009})}\BibitemShut {NoStop}%
\bibitem [{\citenamefont {Zimmermann}\ \emph {et~al.}(2017)\citenamefont
  {Zimmermann}, \citenamefont {Patchkovskii}, \citenamefont {Ivanov},\ and\
  \citenamefont {Eichmann}}]{Zimmermann2017}%
  \BibitemOpen
  \bibfield  {author} {\bibinfo {author} {\bibfnamefont {H.}~\bibnamefont
  {Zimmermann}}, \bibinfo {author} {\bibfnamefont {S.}~\bibnamefont
  {Patchkovskii}}, \bibinfo {author} {\bibfnamefont {M.}~\bibnamefont
  {Ivanov}},\ and\ \bibinfo {author} {\bibfnamefont {U.}~\bibnamefont
  {Eichmann}},\ }\bibfield  {title} {\bibinfo {title} {Unified time and
  frequency picture of ultrafast atomic excitation in strong laser fields},\
  }\href {https://doi.org/10.1103/physrevlett.118.013003} {\bibfield  {journal}
  {\bibinfo  {journal} {Physical Review Letters}\ }\textbf {\bibinfo {volume}
  {118}},\ \bibinfo {pages} {013003} (\bibinfo {year} {2017})}\BibitemShut
  {NoStop}%
\bibitem [{\citenamefont {Freeman}\ \emph {et~al.}(1987)\citenamefont
  {Freeman}, \citenamefont {Bucksbaum}, \citenamefont {Milchberg},
  \citenamefont {Darack}, \citenamefont {Schumacher},\ and\ \citenamefont
  {Geusic}}]{Freeman1987}%
  \BibitemOpen
  \bibfield  {author} {\bibinfo {author} {\bibfnamefont {R.~R.}\ \bibnamefont
  {Freeman}}, \bibinfo {author} {\bibfnamefont {P.~H.}\ \bibnamefont
  {Bucksbaum}}, \bibinfo {author} {\bibfnamefont {H.}~\bibnamefont
  {Milchberg}}, \bibinfo {author} {\bibfnamefont {S.}~\bibnamefont {Darack}},
  \bibinfo {author} {\bibfnamefont {D.}~\bibnamefont {Schumacher}},\ and\
  \bibinfo {author} {\bibfnamefont {M.~E.}\ \bibnamefont {Geusic}},\ }\bibfield
   {title} {\bibinfo {title} {Above-threshold ionization with subpicosecond
  laser pulses},\ }\href {https://doi.org/10.1103/physrevlett.59.1092}
  {\bibfield  {journal} {\bibinfo  {journal} {Physical Review Letters}\
  }\textbf {\bibinfo {volume} {59}},\ \bibinfo {pages} {1092} (\bibinfo {year}
  {1987})}\BibitemShut {NoStop}%
\bibitem [{\citenamefont {Dill}(1973)}]{Dill1973}%
  \BibitemOpen
  \bibfield  {author} {\bibinfo {author} {\bibfnamefont {D.}~\bibnamefont
  {Dill}},\ }\bibfield  {title} {\bibinfo {title} {Resonances in photoelectron
  angular distributions},\ }\href {https://doi.org/10.1103/physreva.7.1976}
  {\bibfield  {journal} {\bibinfo  {journal} {Physical Review A}\ }\textbf
  {\bibinfo {volume} {7}},\ \bibinfo {pages} {1976} (\bibinfo {year}
  {1973})}\BibitemShut {NoStop}%
\bibitem [{\citenamefont {Samson}\ and\ \citenamefont
  {Gardner}(1973)}]{Samson1973}%
  \BibitemOpen
  \bibfield  {author} {\bibinfo {author} {\bibfnamefont {J.~A.~R.}\
  \bibnamefont {Samson}}\ and\ \bibinfo {author} {\bibfnamefont {J.~L.}\
  \bibnamefont {Gardner}},\ }\bibfield  {title} {\bibinfo {title} {Resonances
  in the angular distribution of xenon photoelectrons},\ }\href
  {https://doi.org/10.1103/physrevlett.31.1327} {\bibfield  {journal} {\bibinfo
   {journal} {Physical Review Letters}\ }\textbf {\bibinfo {volume} {31}},\
  \bibinfo {pages} {1327} (\bibinfo {year} {1973})}\BibitemShut {NoStop}%
\bibitem [{\citenamefont {Saloman}(2004)}]{Saloman2004}%
  \BibitemOpen
  \bibfield  {author} {\bibinfo {author} {\bibfnamefont {E.~B.}\ \bibnamefont
  {Saloman}},\ }\bibfield  {title} {\bibinfo {title} {Energy levels and
  observed spectral lines of xenon, {Xe} {I} through {Xe} {LIV}},\ }\href
  {https://doi.org/10.1063/1.1649348} {\bibfield  {journal} {\bibinfo
  {journal} {Journal of Physical and Chemical Reference Data}\ }\textbf
  {\bibinfo {volume} {33}},\ \bibinfo {pages} {765} (\bibinfo {year}
  {2004})}\BibitemShut {NoStop}%
\bibitem [{\citenamefont {Petite}\ \emph {et~al.}(1987)\citenamefont {Petite},
  \citenamefont {Agostini},\ and\ \citenamefont {Yergeau}}]{Petite1987}%
  \BibitemOpen
  \bibfield  {author} {\bibinfo {author} {\bibfnamefont {G.}~\bibnamefont
  {Petite}}, \bibinfo {author} {\bibfnamefont {P.}~\bibnamefont {Agostini}},\
  and\ \bibinfo {author} {\bibfnamefont {F.}~\bibnamefont {Yergeau}},\
  }\bibfield  {title} {\bibinfo {title} {Intensity, pulse width, and
  polarization dependence of above-threshold-ionization electron spectra},\
  }\href {https://doi.org/10.1364/josab.4.000765} {\bibfield  {journal}
  {\bibinfo  {journal} {Journal of the Optical Society of America B}\ }\textbf
  {\bibinfo {volume} {4}},\ \bibinfo {pages} {765} (\bibinfo {year}
  {1987})}\BibitemShut {NoStop}%
\bibitem [{\citenamefont {Carlström}\ \emph {et~al.}(2018)\citenamefont
  {Carlström}, \citenamefont {Mauritsson}, \citenamefont {Schafer},
  \citenamefont {L’Huillier},\ and\ \citenamefont
  {Gisselbrecht}}]{Carlstroem2018}%
  \BibitemOpen
  \bibfield  {author} {\bibinfo {author} {\bibfnamefont {S.}~\bibnamefont
  {Carlström}}, \bibinfo {author} {\bibfnamefont {J.}~\bibnamefont
  {Mauritsson}}, \bibinfo {author} {\bibfnamefont {K.~J.}\ \bibnamefont
  {Schafer}}, \bibinfo {author} {\bibfnamefont {A.}~\bibnamefont
  {L’Huillier}},\ and\ \bibinfo {author} {\bibfnamefont {M.}~\bibnamefont
  {Gisselbrecht}},\ }\bibfield  {title} {\bibinfo {title} {Quantum coherence in
  photo-ionisation with tailored {XUV} pulses},\ }\href
  {https://doi.org/10.1088/1361-6455/aa96e7} {\bibfield  {journal} {\bibinfo
  {journal} {Journal of Physics B: Atomic, Molecular and Optical Physics}\
  }\textbf {\bibinfo {volume} {51}},\ \bibinfo {pages} {015201} (\bibinfo
  {year} {2018})}\BibitemShut {NoStop}%
\bibitem [{\citenamefont {Varshalovich}(1988)}]{Varshalovich1988}%
  \BibitemOpen
  \bibfield  {author} {\bibinfo {author} {\bibfnamefont {D.~A.}\ \bibnamefont
  {Varshalovich}},\ }\href@noop {} {\emph {\bibinfo {title} {Quantum Theory of
  Angular Momentum: Irreducible Tensors, Spherical Harmonics, Vector Coupling
  Coefficients, 3nj Symbols}}}\ (\bibinfo  {publisher} {World Scientific Pub},\
  \bibinfo {address} {Singapore Teaneck, NJ, USA},\ \bibinfo {year}
  {1988})\BibitemShut {NoStop}%
\bibitem [{\citenamefont {Adler}\ and\ \citenamefont
  {Piran}(1984)}]{Adler1984}%
  \BibitemOpen
  \bibfield  {author} {\bibinfo {author} {\bibfnamefont {S.~L.}\ \bibnamefont
  {Adler}}\ and\ \bibinfo {author} {\bibfnamefont {T.}~\bibnamefont {Piran}},\
  }\bibfield  {title} {\bibinfo {title} {Relaxation methods for gauge field
  equilibrium equations},\ }\href {https://doi.org/10.1103/revmodphys.56.1}
  {\bibfield  {journal} {\bibinfo  {journal} {Reviews of Modern Physics}\
  }\textbf {\bibinfo {volume} {56}},\ \bibinfo {pages} {1} (\bibinfo {year}
  {1984})}\BibitemShut {NoStop}%
\bibitem [{\citenamefont {Krause}\ and\ \citenamefont
  {Schafer}(1999)}]{Krause1999TJoPCA}%
  \BibitemOpen
  \bibfield  {author} {\bibinfo {author} {\bibfnamefont {J.~L.}\ \bibnamefont
  {Krause}}\ and\ \bibinfo {author} {\bibfnamefont {K.~J.}\ \bibnamefont
  {Schafer}},\ }\bibfield  {title} {\bibinfo {title} {Control of {THz} emission
  from {S}tark wave packets},\ }\href {https://doi.org/10.1021/jp992144}
  {\bibfield  {journal} {\bibinfo  {journal} {The Journal of Physical Chemistry
  A}\ }\textbf {\bibinfo {volume} {103}},\ \bibinfo {pages} {10118} (\bibinfo
  {year} {1999})}\BibitemShut {NoStop}%
\bibitem [{\citenamefont {Garstang}(1964)}]{Garstang1964}%
  \BibitemOpen
  \bibfield  {author} {\bibinfo {author} {\bibfnamefont {R.}~\bibnamefont
  {Garstang}},\ }\bibfield  {title} {\bibinfo {title} {Transition probabilities
  of forbidden lines},\ }\href {https://doi.org/10.6028/jres.068a.004}
  {\bibfield  {journal} {\bibinfo  {journal} {Journal of Research of the
  National Bureau of Standards Section A: Physics and Chemistry}\ }\textbf
  {\bibinfo {volume} {68A}},\ \bibinfo {pages} {61} (\bibinfo {year}
  {1964})}\BibitemShut {NoStop}%
\bibitem [{\citenamefont {Nandy}\ and\ \citenamefont
  {Sahoo}(2015)}]{Nandy2015}%
  \BibitemOpen
  \bibfield  {author} {\bibinfo {author} {\bibfnamefont {D.~K.}\ \bibnamefont
  {Nandy}}\ and\ \bibinfo {author} {\bibfnamefont {B.~K.}\ \bibnamefont
  {Sahoo}},\ }\bibfield  {title} {\bibinfo {title} {Forbidden transition
  properties in the ground-state configurations of singly ionized noble gas
  atoms for stellar and interstellar media},\ }\href
  {https://doi.org/10.1093/mnras/stv683} {\bibfield  {journal} {\bibinfo
  {journal} {Monthly Notices of the Royal Astronomical Society}\ }\textbf
  {\bibinfo {volume} {450}},\ \bibinfo {pages} {1012} (\bibinfo {year}
  {2015})}\BibitemShut {NoStop}%
\bibitem [{\citenamefont {Dolg}\ and\ \citenamefont {Cao}(2011)}]{Dolg2011}%
  \BibitemOpen
  \bibfield  {author} {\bibinfo {author} {\bibfnamefont {M.}~\bibnamefont
  {Dolg}}\ and\ \bibinfo {author} {\bibfnamefont {X.}~\bibnamefont {Cao}},\
  }\bibfield  {title} {\bibinfo {title} {Relativistic pseudopotentials: Their
  development and scope of applications},\ }\href
  {https://doi.org/10.1021/cr2001383} {\bibfield  {journal} {\bibinfo
  {journal} {Chemical Reviews}\ }\textbf {\bibinfo {volume} {112}},\ \bibinfo
  {pages} {403} (\bibinfo {year} {2011})}\BibitemShut {NoStop}%
\bibitem [{\citenamefont {Manolopoulos}(2002)}]{Manolopoulos2002}%
  \BibitemOpen
  \bibfield  {author} {\bibinfo {author} {\bibfnamefont {D.~E.}\ \bibnamefont
  {Manolopoulos}},\ }\bibfield  {title} {\bibinfo {title} {Derivation and
  reflection properties of a transmission-free absorbing potential},\ }\href
  {https://doi.org/10.1063/1.1517042} {\bibfield  {journal} {\bibinfo
  {journal} {J. Chem. Phys.}\ }\textbf {\bibinfo {volume} {117}},\ \bibinfo
  {pages} {9552} (\bibinfo {year} {2002})}\BibitemShut {NoStop}%
\bibitem [{\citenamefont {Patchkovskii}\ and\ \citenamefont
  {Muller}(2016)}]{Patchkovskii2016}%
  \BibitemOpen
  \bibfield  {author} {\bibinfo {author} {\bibfnamefont {S.}~\bibnamefont
  {Patchkovskii}}\ and\ \bibinfo {author} {\bibfnamefont {H.}~\bibnamefont
  {Muller}},\ }\bibfield  {title} {\bibinfo {title} {Simple, accurate, and
  efficient implementation of 1-electron atomic time-dependent
  {S}chr{\"o}dinger equation in spherical coordinates},\ }\href
  {https://doi.org/10.1016/j.cpc.2015.10.014} {\bibfield  {journal} {\bibinfo
  {journal} {Computer Physics Communications}\ }\textbf {\bibinfo {volume}
  {199}},\ \bibinfo {pages} {153} (\bibinfo {year} {2016})}\BibitemShut
  {NoStop}%
\bibitem [{\citenamefont {Fritsch}\ and\ \citenamefont
  {Lin}(1991)}]{Fritsch1991}%
  \BibitemOpen
  \bibfield  {author} {\bibinfo {author} {\bibfnamefont {W.}~\bibnamefont
  {Fritsch}}\ and\ \bibinfo {author} {\bibfnamefont {C.-D.}\ \bibnamefont
  {Lin}},\ }\bibfield  {title} {\bibinfo {title} {The semiclassical
  close-coupling description of atomic collisions: Recent developments and
  results},\ }\href {https://doi.org/10.1016/0370-1573(91)90008-a} {\bibfield
  {journal} {\bibinfo  {journal} {Physics Reports}\ }\textbf {\bibinfo {volume}
  {202}},\ \bibinfo {pages} {1} (\bibinfo {year} {1991})}\BibitemShut {NoStop}%
\bibitem [{\citenamefont {Jankowiak}\ \emph {et~al.}(1993)\citenamefont
  {Jankowiak}, \citenamefont {Hayes},\ and\ \citenamefont
  {Small}}]{Jankowiak1993}%
  \BibitemOpen
  \bibfield  {author} {\bibinfo {author} {\bibfnamefont {R.}~\bibnamefont
  {Jankowiak}}, \bibinfo {author} {\bibfnamefont {J.~M.}\ \bibnamefont
  {Hayes}},\ and\ \bibinfo {author} {\bibfnamefont {G.~J.}\ \bibnamefont
  {Small}},\ }\bibfield  {title} {\bibinfo {title} {Spectral hole-burning
  spectroscopy in amorphous molecular solids and proteins},\ }\href
  {https://doi.org/10.1021/cr00020a005} {\bibfield  {journal} {\bibinfo
  {journal} {Chemical Reviews}\ }\textbf {\bibinfo {volume} {93}},\ \bibinfo
  {pages} {1471} (\bibinfo {year} {1993})}\BibitemShut {NoStop}%
\bibitem [{\citenamefont {Goll}\ \emph {et~al.}(2006)\citenamefont {Goll},
  \citenamefont {Wunner},\ and\ \citenamefont {Saenz}}]{Goll2006}%
  \BibitemOpen
  \bibfield  {author} {\bibinfo {author} {\bibfnamefont {E.}~\bibnamefont
  {Goll}}, \bibinfo {author} {\bibfnamefont {G.}~\bibnamefont {Wunner}},\ and\
  \bibinfo {author} {\bibfnamefont {A.}~\bibnamefont {Saenz}},\ }\bibfield
  {title} {\bibinfo {title} {Formation of ground-state vibrational wave packets
  in intense ultrashort laser pulses},\ }\href
  {https://doi.org/10.1103/physrevlett.97.103003} {\bibfield  {journal}
  {\bibinfo  {journal} {Physical Review Letters}\ }\textbf {\bibinfo {volume}
  {97}},\ \bibinfo {pages} {103003} (\bibinfo {year} {2006})}\BibitemShut
  {NoStop}%
\end{thebibliography}%

\end{document}